\documentstyle[12pt]{article}
\input{psfig.sty}

\begin{document}

\begin{titlepage}

\begin{flushright}
RAL-TR-98-022
\end{flushright}

\baselineskip 24pt

\begin{center}

{\Large {\bf Features of Quark and Lepton Mixing from  Differential Geometry 
of Curves on Surfaces}}\\

\vspace{.2cm}

\baselineskip 12pt
{\large Jos\'e BORDES}\\
bordes\,@\,evalvx.ific.uv.es\\
{\it Dept. Fisica Teorica, Univ. de Valencia,\\
  c. Dr. Moliner 50, E-46100 Burjassot (Valencia), Spain}\\
{\large CHAN Hong-Mo}\\
chanhm\,@\,v2.rl.ac.uk\\
{\it Rutherford Appleton Laboratory,\\
  Chilton, Didcot, Oxon, OX11 0QX, United Kingdom}\\
{\large Jakov PFAUDLER}\\
jakov\,@\,thphys.ox.ac.uk\\
{\it Dept. of Physics, Theoretical Physics, University of Oxford,\\
  1 Keble Road, Oxford, OX1 3NP, United Kingdom}\\
{\large TSOU Sheung Tsun}\\
tsou\,@\,maths.ox.ac.uk\\
{\it Mathematical Institute, University of Oxford,\\
  24-29 St. Giles', Oxford, OX1 3LB, United Kingdom}
\end{center}

\baselineskip 12pt

\begin{abstract}

It is noted that the CKM matrix elements for both quarks and leptons
as conceived 
in the Dualized Standard Model (DSM) can be interpreted as direction
cosines obtained by moving the Darboux trihedron (a 3-frame)  
along a trajectory on a sphere traced out through changing energy 
scales by a 3-vector factorized from the mass matrix.  From the `Darboux'
analogues of the well-known Serret--Frenet formulae for space curves, it is
seen that the corner elements ($V_{ub}, V_{td}$ for quarks, and $U_{e3}, 
U_{\tau 1}$ for leptons) are associated with the (geodesic) torsion, while 
the other off-diagonal elements ($V_{us}, V_{cd}$ and $V_{cb}, V_{ts}$ for 
quarks, and $U_{e2}, U_{\mu 1}$ and $U_{\mu 3}, U_{\tau 2}$ for leptons) 
with the (respectively geodesic and normal) curvatures of the 
trajectory.  From this it follows that (i) the corner elements in both
matrices are 
much smaller than the other elements, (ii) the $U_{\mu 3}, U_{\tau 2}$ 
elements for the lepton CKM matrix are much larger than their counterparts 
in the quark matrix.  Both these conclusions are strongly borne out 
by experiment, for quarks in hadron decays and for leptons in neutrino 
oscillations, and by previous explicit calculations within the DSM scheme.  

\end{abstract}

\end{titlepage}

\clearpage

\baselineskip 14pt

The Cabibbo-Kobayashi-Moskawa matrix which gives the relative orientation
in generation space between the state vectors of the three physical states
of $U$-type and $D$-type quarks has now been measured to a fair accuracy
by experiment.  In the latest databook \cite{databook}, it is given as:
\begin{equation}
|CKM|_{quark} = \left( \begin{array}{lll} 
       0.9745 - 0.9760 & 0.217 - 0.224 & 0.0018 - 0.0045 \\
       0.217 - 0.224 & 0.9737 - 0.9753 & 0.036 - 0.042 \\
       0.004 - 0.013 & 0.035 - 0.042 & 0.9991 - 0.9994 \end{array} \right),
\label{quarkCKM}
\end{equation}
where the modulus sign on the left-hand side means that we are giving
only the absolute values of the entries.

A similar matrix can in principle be defined between charged leptons and
neutrinos, here referred to as the leptonic CKM matrix, for which the
experimental information is still rather fragmentary in comparison, but some 
features of which are beginning to emerge.  First, from the so-called muon 
anomaly observed in atmospheric neutrinos \cite{Kamioka,Superk,IMB,Soudan},
one obtains that the matrix element relating the muon to the heaviest 
neutrino $\nu_3$, namely $|U_{\mu 3}|$ is rather large, say roughly 
$0.45 - 0.85$, according to a recent analysis \cite{Giunkimno}.  Secondly, 
from the absence of any oscillation effects in reactor experiments such as 
CHOOZ \cite{Chooz}, one obtains, at least for the squared mass difference
$\Delta m_{23}^2$ of the two heaviest neutrinos being in the range 
$10^{-2} - 10^{-3}$ eV$^2$ as favoured by the atmospheric neutrino
data, that the element relating the electron to $\nu_3$, namely $|U_{e 3}|$,
is small, say roughly less than 0.15 \cite{Giunkimno}.  Thirdly, for the
solar neutrino puzzle \cite{solarrev}, one is offered several solutions,
either (i) the so-called long wave-length oscillation (LWO) solution which
wants $|U_{e 2}| \sim 0.4 - 0.7$ \cite{Bargphil,Krascov}, or (ii) the 
so-called MSW solution making use of the Mikheyev-Smirnov-Wolfenstein 
\cite{Wolfenstein,Mikhenov} mechanism which wants either $|U_{e 2}| \sim 
0.4 - 0.6$ (large angle solution) or $|U_{e 2}| \sim 0.03 - 0.05$ (small 
angle solution) \cite{MSWfit}.  If we choose either the LWO solution or
the large angle MSW solution for which there seems to be some preference, 
we obtain for the leptonic CKM matrix the following tentative arrangement:
\begin{equation}
|CKM|_{lepton} = \left( \begin{array}{ccc}
       \ast & 0.4 - 0.7 & 0.0 - 0.15 \\
       \ast & \ast & 0.45 - 0.85 \\
       \ast & \ast & \ast \end{array} \right).
\label{leptonCKM}
\end{equation}
In as much as the CKM matrix is a rotation matrix near to the identity, 
$|CKM|$ is roughly symmetric.  And given that we are at present ignoring 
the CP-violating phase, it contains only 3 independent real parameters,
so that from the bounds of the 3 elements specified above rough bounds 
for the other elements can be calculated, but we shall not need them for 
our present discussion.

Comparing the two matrices (\ref{quarkCKM}) and (\ref{leptonCKM}), one 
notices the following salient features.  First, the corner elements 13 
and 31 are in both cases much smaller than the rest.  Second, the 23 
element of the lepton CKM matrix is much larger than its counterpart
in the quark CKM matrix.  Third, the 12 (21) elements for both matrices
are sizeable and of roughly the same magnitude.  Now these features are 
highly significant in physical terms, for the quark case in explaining 
the branching ratios of various hadronic decays, and for the lepton case, 
as outlined above, in explaining certain effects in neutrino oscillations.  
It would thus be important to understand how these features arise in any 
attempt at modelling quark and lepton mixing.

Now we have recently suggested a scheme based on the dual properties of
the Standard Model (DSM) \cite{Chantsou} which purports to explain the 
existence of exactly three generations as spontaneously broken dual colour 
and affords at the same time a means for evaluating perturbatively the 
CKM matrix for both the quark \cite{OurCKM} and the lepton \cite{Ournuos} 
case.  Calculations to 1-loop level have already been performed giving 
quite satisfactory answers which have the empirical features emphasized 
in the above paragraph.  It would thus be interesting to find the theoretical 
reason within the scheme why these particular features should emerge.

Let us recall some basic properties of the fermion mass matrix in the DSM 
scheme.  By virtue of the manner in which the generation (dual colour) 
symmetry is broken, as suggested by the Higgs fields occurring naturally 
in the theory, the fermion mass matrix in the DSM is factorizable at 
tree-level, 
and even remains so after loop corrections \cite{Chantsou}.  Thus,
\begin{equation}
m' = m_T \left( \begin{array}{c} x' \\ y' \\ z' \end{array} \right) (x',y',z').
\label{massmat}
\end{equation}
Apart from the mass scale $m_T$ which may be taken to be the mass of the
highest state of the fermion-type under consideration, the whole
information of the mass 
matrix is encoded in the vector $(x',y',z')$.  This vector rotates with
changing energy scales, thus tracing out a trajectory in generation space, 
and in the approximation where one neglects the scale-dependence of the length
of $(x',y',z')$ which one cannot compute at present, one can normalize
the vector thus:
\begin{equation}
x'^2 + y'^2 + z'^2 = 1,
\label{xyznorm}
\end{equation}
so that the trajectory lies on the unit sphere.  Further, without loss
of generality we can take $x' \geq y' \geq z'$.

\begin{figure}[htb]
\vspace{-5cm}
\centerline{\psfig{figure=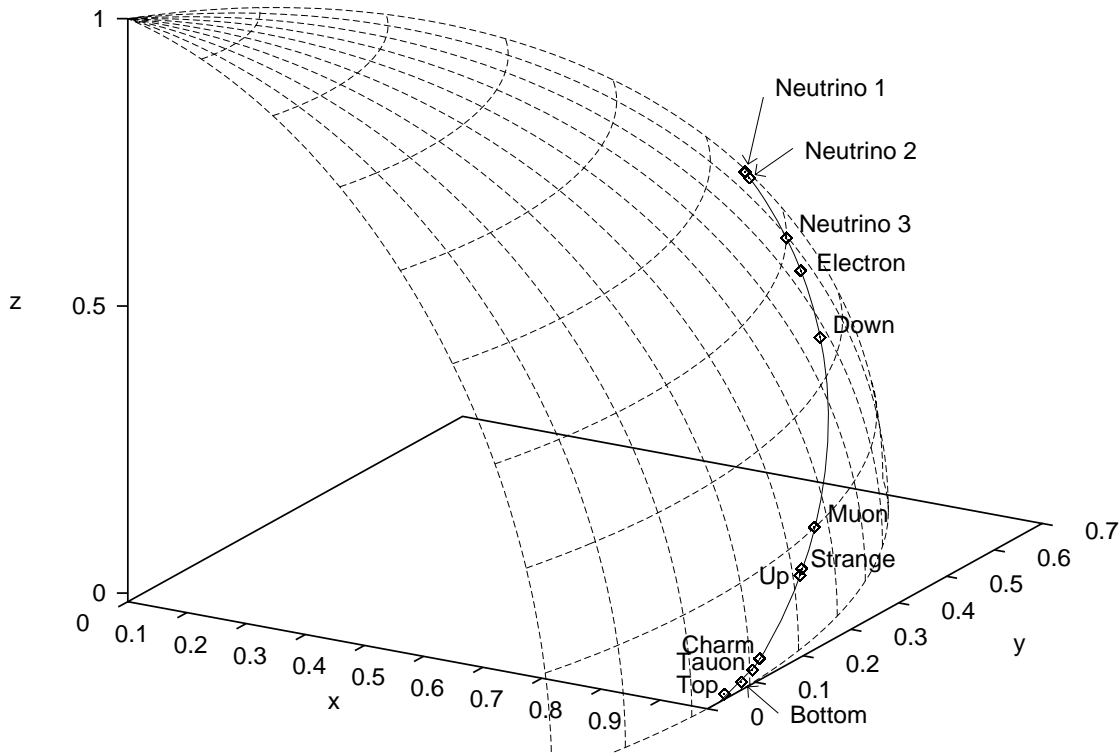,width=0.85\textwidth}}
\vspace{0cm} 
\caption{The trajectory traced out by $(x',y',z')$ and the locations on
it of the 12 physical fermion states.}
\label{runtraj}
\end{figure}

Although the trajectory traced out by $(x',y',z')$ can in principle be 
different for different fermion-types it was found in \cite{OurCKM} by
fitting experiment that, to a very good accuracy, all fermion-types run 
on the same trajectory with the same speed with respect to the energy scale
$\mu$, only differing by the locations of the physical states.  The actual 
trajectory obtained in \cite{OurCKM} together with the locations on it 
of the 12 physical states \cite{OurCKM,Ournuos} are shown in Figure
\ref{runtraj}.

We shall not detail here the criterion used in the DSM for determining 
the location, namely the mass, of the lower generation states, since it
is of no great relevance for our present discussion, but refer the reader
to \cite{OurCKM}.  What matters for us here is how the state vectors in 
generation (i.e. dual colour) space of the physical states are defined.

First, for each fermion-type (i.e.\ whether $U$ for $U$-type quarks,
$D$ for $D$-type
quarks, $L$ for charged leptons or $N$ for neutrinos), the highest generation
state (i.e.\ $t, b, \tau, \nu_3$ in the above order) is defined as that
eigenstate $|{\bf v}_1 \rangle$ of the mass matrix $m'$ in (\ref{massmat}) 
with non-zero eigenvalue, taken at the scale equal to the mass of that 
generation, namely at $\mu = m_t, m_b, m_\tau, m_{\nu_3}$ respectively.
It is easily seen that in Figure \ref{runtraj} these are just the radial 
vectors $(x',y',z')$ themselves taken at the appropriate locations.

\begin{figure}[htb]
\vspace{-2cm}
\centerline{\psfig{figure=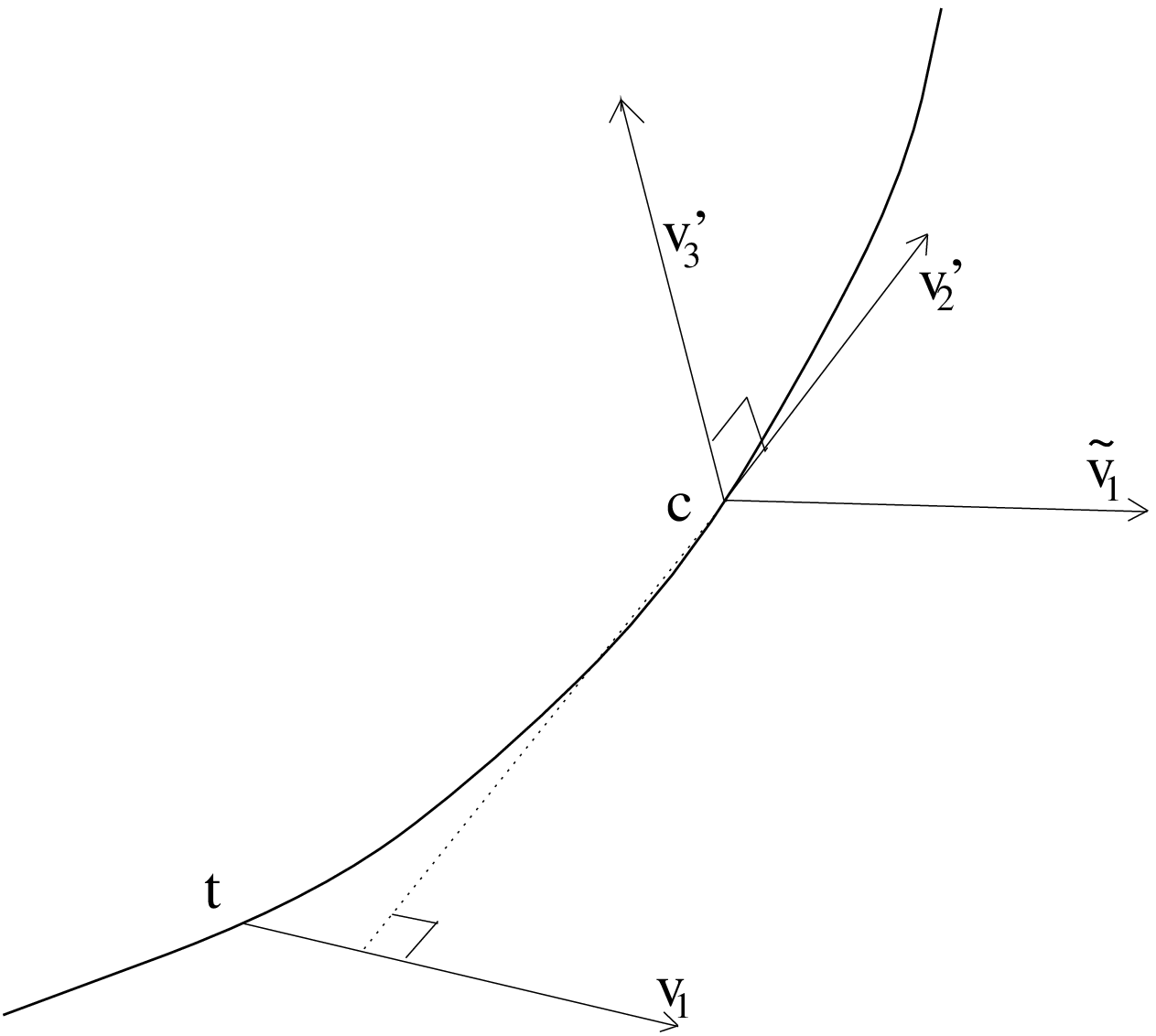,width=0.6\textwidth}}
\vspace{0cm} 
\caption{The state vectors of the 3 physical states belonging to the
3 generations of each fermion-type.}
\label{v1v2v3}
\end{figure}

Secondly, the state vector of the second generation which we called 
$|{\bf v}'_2 \rangle$ in \cite{OurCKM} is defined again as the eigenvector 
with non-zero eigenvalue but now of the submatrix of $m'$ in the 2-dimensional
subspace orthogonal to $|{\bf v}_1 \rangle$ evaluated at the scale equal
to the second generation mass.  At the location on the trajectory occupied
by the second generation, the vector $(x',y',z')$ would have rotated to a
different direction, say $|{\tilde {\bf v}}_1 \rangle$, as illustrated in 
Figure \ref{v1v2v3}.  The vector $|{\bf v}'_2 \rangle$ is the vector 
orthogonal to $|{\bf v}_1 \rangle$ which soaks up all the `leakage' 
from this rotation.  It is easy to see then that $|{\bf v}'_2 \rangle$ 
lies on the plane 
formed by $|{\bf v}_1 \rangle$ and $|{\tilde {\bf v}}_1 \rangle$ as shown 
in Figure \ref{v1v2v3}.
Obviously, the state vector $|{\bf v}'_3 \rangle$ of the lowest generation 
is then, as also shown in the figure, the vector orthogonal to both 
$|{\bf v}_1 \rangle$ and $|{\bf v}'_2 \rangle$.

Imagine now that the locations of the highest two generations are rather
close together, which is indeed the case for all the 4 fermion-types shown
in Figure \ref{runtraj}.  In that case, the vector $|{\bf v}'_2 \rangle$
approaches the tangent vector of the trajectory at $(x',y',z')$.  In other
words, in this approximation, the 3 state vectors for the 3 generations
coincide with the triad formed by the radial vector, the tangent vector to 
the trajectory, and the normal vector to both.

\begin{figure}[htb]
\vspace{-2cm}
\centerline{\psfig{figure=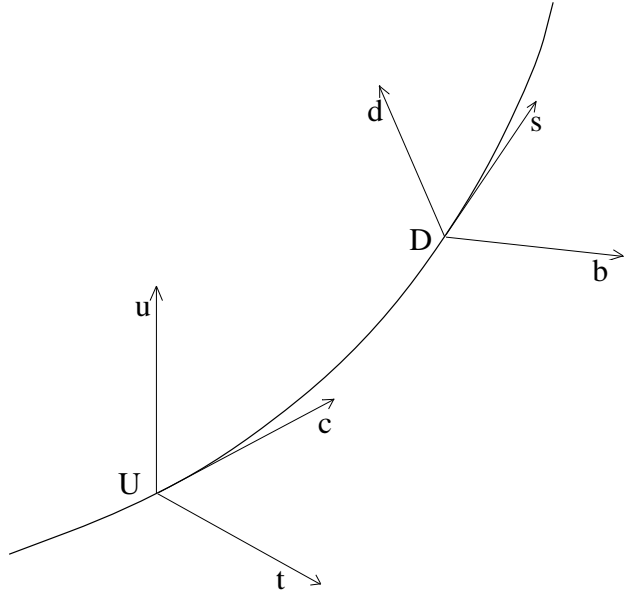,width=0.6\textwidth}}
\vspace{0cm} 
\caption{Two triads of state vectors for two fermion-types
transported along a common trajectory.}
\label{2triads}
\end{figure}

Next, still in this approximation, consider the triad of state vectors for a 
pair of fermion-types, i.e. either $U$- with $D$-type quarks, or the charged 
leptons $L$ with the neutrinos $N$.  At the location (of the combined two 
highest generations) of each fermion-type, we have associated a triad of 
state vectors as depicted in Figure \ref{2triads}.  The triad at one location 
is obtained just by transporting the triad from the other 
location along the trajectory on the sphere.
Now the CKM matrix is by definition the matrix giving the relative
orientation of the triad of physical state vectors between the $U$-type
and $D$-type quarks, or in the leptonic version between the charge leptons
and the neutrinos.  In view of our observations above, we conclude that
it can be written as:
\begin{equation}
CKM = P_s \exp  \int_U^D ds  A (s),
\label{Diracphase}
\end{equation}
similar in form to the Dirac phase factor in gauge theory language,
where $s(\mu)$ is  the distance on 
the trajectory, $\mu$ is the energy scale, and 
$A$ is some rotation matrix yet to be specified.  Notice 
that to 1-loop order, the distance $s$ travelled per unit change in the 
energy scale $\mu$ is proportional to the Yukawa coupling strength squared 
$\rho^2$, which therefore plays the role of the gauge coupling in the 
usual formula for the Dirac phase factor in gauge theory.

To specify the matrix $A$, we let the two locations corresponding to 
the two fermion-types approach each other on the trajectory and evaluate the 
CKM matrix to first order in the distance $\Delta s$ between the two locations.
This can easily be done using some elementary results from the differential 
geometry of curves on surfaces \cite{Docarmo}.  Let $\Gamma$ be a curve
parametrized by arc-length $s$ lying on a surface $S$, in our case the
unit sphere.  At every point on the curve $\Gamma$, we have {\bf N} the
unit normal to the surface, {\bf T} the unit tangent to the curve, and we
can define a third unit vector {\bf B}={\bf N} $\wedge$ {\bf T} normal
to both.  These form what is sometimes called the Darboux trihedron and 
are actually exactly the triad we are interested in, since for the sphere 
the radial and normal vectors coincide.  Differentiating these vectors
with respect to arc-length we get the following relations which we may
call the Serret--Frenet--Darboux formulae in analogy to the usual
Serret--Frenet formulae for space curves:
\begin{eqnarray}
{\bf N}' & = & -\kappa_n {\bf T} - \tau_g {\bf B}, \nonumber\\
{\bf T}' & = &  \kappa_g {\bf B} + \kappa_n {\bf N}, \nonumber \\
{\bf B}' & = & -\tau_g {\bf N} - \kappa_g {\bf T}.
\label{SFDarboux}
\end{eqnarray}
Here $\kappa_g$ is the geodesic curvature, $\kappa_n$ the normal curvature, 
and $\tau_g$ the geodesic torsion of the curve $\Gamma$ on $S$. Recall 
that the Serret--Frenet formula for ${\bf T}'$ defines both the curvature 
$\kappa$ of the curve and its normal {\bf n} at the given point:
\begin{equation}
{\bf T}'=\kappa {\bf n},
\label{Tprime}
\end{equation}
and that, being normal to ${\bf T}$, the curvature vector $\kappa {\bf n}$
lies on the ${\bf N}-{\bf B}$ plane.  The second equation in (\ref{SFDarboux})
gives its decomposition along ${\bf B}$ and ${\bf N}$.  Arranged in the 
conventional order for physicists in the form of a CKM matrix\footnote{Notice 
that it is conventional to label the rows and columns of the CKM matrix in 
order of increasing mass, which is opposite to that adopted above, following 
\cite{OurCKM}, for labelling the state vectors $|{\bf v}'_i \rangle$ of the 
three generations.}, the formulae in (\ref{SFDarboux}) read as:
\begin{equation}
CKM \sim \left( \begin{array}{ccc}
       1 & -\kappa_g \Delta s & -\tau_g \Delta s \\
       \kappa_g \Delta s & 1 & \kappa_n \Delta s \\
       \tau_g \Delta s & -\kappa_n \Delta s & 1 \end{array} \right),
\label{approxCKM}
\end{equation}
to first order in $\Delta s$, this being the distance between the locations
for the two fermion-types.  This gives the matrix $A$ in
(\ref{Diracphase}) as:
\begin{equation}
 A = \left( \begin{array}{ccc} 
          0 & -\kappa_g & -\tau_g \\
          \kappa_g & 0 & \kappa_n \\
          \tau_g & -\kappa_n & 0 \end{array} \right).
\label{Ai}
\end{equation}

Suppose now that the separation in locations between the two-fermion types 
on the trajectory is actually small as is indeed the case for the $U$- 
and $D$-type quarks, as seen in Figure \ref{runtraj}.  The CKM matrix is
then to a good approximation given just by (\ref{approxCKM}) above.  We 
notice first that the off-diagonal corner elements of the matrix are 
proportional to the geodesic torsion $\tau_g$ while the other off-diagonal
elements are proportional to the curvatures $\kappa_g$ and $\kappa_n$.  
For the particular case of a curve on a sphere, which is our case, the 
geodesic torsion $\tau_g$ vanishes \cite{Docarmo}, so that the corner elements 
of the CKM matrix are of second order in $\Delta s$ and therefore relatively
small.  Comparing this with the empirical CKM matrix for quarks quoted in 
(\ref{quarkCKM}), one sees that this prediction is very clearly borne out.  
Secondly, the elements $V_{cb}$ and $V_{ts}$ are proportional to the normal
curvature $\kappa_n$ which for the sphere is constant, independent either 
of the trajectory or of the location.  It follows then from (\ref{approxCKM})
that their values are approximately given just by the distance, measured in 
units of the radius of the sphere, between the locations of $t$ and $b$ 
on the trajectory.  From Figure \ref{runtraj}, one estimates this to be of 
order of a few percent.  Again, the prediction is seen to be borne out 
in (\ref{quarkCKM}).  Both these special features are also what one finds in 
the explicit calculation carried out in \cite{OurCKM}.  As for the remaining
off-diagonal elements $V_{us}$ and $V_{cd}$, namely the Cabibbo angle,
it is in a sense a little special.  Being proportional to the geodesic 
curvature $\kappa_g$ which depends both on the location and on the 
trajectory, it can be sizeable even for small separation.  This is seen 
to be indeed the case in the empirical matrix (\ref{quarkCKM}) and agrees
with the fact that in the calculation \cite{OurCKM}, the Cabibbo angle 
can be fitted to the experimental value by choosing the trajectory.

What would happen in the case of the leptons where the separation between
the two fermion-types, i.e.\ between $\tau$ and $\nu_3$, is not so small, as 
seen in Figure \ref{runtraj}?  We shall need in that case to evaluate the 
`Dirac phase factor' (\ref{Diracphase}) above, as was
done numerically in \cite{Ournuos}.  However, even without an explicit 
calculation we can already draw some qualitative conclusions.  First, the
corner element $U_{e 3}$ being of second order in the distance travelled
will remain small.  This checks with the empirical estimates quoted in 
(\ref{leptonCKM}) and with the calculation \cite{Ournuos}.  Secondly, 
the normal curvature $\kappa_n$ being constant all along the trajectory, 
its contribution to the element $U_{\mu 3}$ will accumulate over the larger 
distance here and become much larger than its counterpart $V_{cb}$ in the 
quark case.  In fact, if one just crudely takes the formula (\ref{approxCKM})
and assumes that the element is still roughly proportional to the distance
travelled from $\tau$ to $\nu_3$, one obtains from Figure \ref{runtraj}
that $U_{\mu 3}$ would be some 20 times larger than $V_{cb}$ in the quark
matrix, which estimate is seen already to be in fair agreement both with 
the empirical matrices in (\ref{quarkCKM}) and (\ref{leptonCKM}), and with 
the explicit calculations in \cite{OurCKM,Ournuos}.  Thirdly, in contrast, 
the remaining off-diagonal element $U_{e 2}$ depends on the geodesic 
curvature $\kappa_g$ which varies with the location.  Its value therefore
cannot be estimated from just the separation between the locations of the
two fermion-types.  There is thus no reason to expect a value for it very 
different in magnitude from the values of $V_{us}$ or $V_{cd}$ in the quark 
case, nor are the values as seen in (\ref{quarkCKM}) and (\ref{leptonCKM})
very different empirically.

There is a further qualitative conclusion that one can draw from the 
preceding analysis about the CKM matrix which, however, cannot be checked 
against experiment but only against the calculation in \cite{OurCKM}.  
As mentioned above, the normal curvature $\kappa_n$ on a sphere does not 
depend on the trajectory.  Hence, the elements $V_{cb}$ and $V_{ts}$, 
being to first order in $\Delta s$ proportional to the normal curvature 
$\kappa_n$, is also expected to be approximately independent of the 
trajectory.  We then see why we were able, when fitting the CKM matrix 
in \cite{OurCKM}, first to fit the elements $V_{cb}$ and $V_{ts}$ by 
adjusting the Higgs vev $y$, and then to fit the rest by adjusting the 
other vev $z$, for the first two elements will remain roughly constant 
in changing $z$ which has the effect mainly of changing the trajectory.

The above analysis was done in the approximations (a) that the locations 
on the trajectory of the two highest states of each fermion-type are
close together, and (b) that the vector $(x',y',z')$ has a normalization
independent of the energy scale $\mu$ and hence moves on a sphere.  However, 
the qualitative features deduced above for the quark and lepton CKM matrices
will still remain even if one relaxes these approximations so long as the 
scale-dependence of the normalization of $(x',y',z')$ is not too strong.  
In any case, the CKM matrix can always be regarded as a `Dirac phase factor' 
of the form (\ref{Diracphase}), only in general the 
matrix $A$ appearing there will be more involved.

Apart from the elegance in giving us a geometrical significance for the 
CKM matrix, and a qualitative understanding of the values of its elements
in simple geometrical terms, the above observations have the practical 
value, we think, in giving us also a nontrivial check on the basic 
structure of the DSM scheme.  The fact that the three off-diagonal
elements 13, 23, and 12 are predicted to have quite different salient
features all agreeing with experiment, depends crucially on the assignment
in the scheme of the highest, middle and lowest generation states to 
respectively the radial vector, the tangent vector and the vector
normal to both, on the trajectory. 
And this trajectory is traced out by the vector $(x',y',z')$, which is 
in turn obtained from a factorizable mass matrix.  Quite a number of 
structural details of the scheme are thus involved in yielding these
predictions.  Further, the fact that rough values of the empirical 
matrix elements can already be deduced just from the Darboux--Serret--Frenet 
formulae (\ref{SFDarboux}) makes the agreement with experiment
obtained in \cite{OurCKM} 
and \cite{Ournuos} much less likely to be mere accidents of the calculations.

\vspace{.3cm}

\noindent{\large {\bf Acknowledgement}}\\

One of us (JB) acknowledges support from the Spanish Government on contract
no. CICYT AEN 97-1718, while another (JP) is grateful to the Studienstiftung
d.d.\ Volkes and the Burton Senior Scholarship of Oriel College, Oxford for
financial support.

\end{document}